\documentstyle[bbm,amsfonts,epsfig,12pt,here]{article}

\newcommand {\eqref} [1] {(\ref {#1})}
\newcommand {\slsh} [1] {\not{\hbox{\kern-2pt${#1}$}}}

\newcommand{\drawsquare}[2]{\hbox{%
\rule{#2pt}{#1pt}\hskip-#2pt
\rule{#1pt}{#2pt}\hskip-#1pt
\rule[#1pt]{#1pt}{#2pt}}\rule[#1pt]{#2pt}{#2pt}\hskip-#2pt
\rule{#2pt}{#1pt}}

\newcommand{\Yfund}{\raisebox{-.5pt}{\drawsquare{6.5}{0.4}}}

\def\drawbox#1#2{\hrule height#2pt
        \hbox{\vrule width#2pt height#1pt \kern#1pt
              \vrule width#2pt}
              \hrule height#2pt}

\def\Asym#1#2{\vcenter{\vbox{\drawbox{#1}{#2}
              \kern-#2pt       
              \drawbox{#1}{#2}}}}

\def\c3z3{{\mathbb C}^3/({\mathbb Z}_3 \otimes{\mathbb Z}_3)}
\def\ckk{{\mathbb C}^3/({\mathbb Z}_k \otimes{\mathbb Z}_{k'})}
\def\cgamma{{\mathbb C}^3/\Gamma}
\newcommand {\beq} {\begin{equation}}
\newcommand {\eeq} {\end{equation}}
 \newcommand {\ber}{\begin{eqnarray*}}
 \newcommand {\eer} {\end{eqnarray*}}
\newcommand {\bea}{\begin{eqnarray}}
 \newcommand {\eea} {\end{eqnarray}}

\begin{document}
\begin{titlepage}
\begin{flushright}{CERN-PH-TH/2004-066
}
\end{flushright}
\vskip 1cm

\centerline{{\Large \bf WITTEN--VENEZIANO from GREEN--SCHWARZ}}
\vskip 1cm
\centerline{\large Adi Armoni}
\vskip 0.1cm
\centerline{\small e-mail: adi.armoni@cern.ch}
\centerline{\small http://armoni.home.cern.ch/armoni/}
\vskip 0.5cm
\centerline{\em Department of Physics, Theory Division, CERN}
\centerline{\em CH-1211 Geneva 23, Switzerland}
\vskip 1cm

\begin{abstract}

We consider the $U(1)$ problem within the AdS/CFT framework. We explain
how the Witten--Veneziano formula for the $\eta '$ mass
is related to a generalized Green--Schwarz mechanism. The closed string mode,
that cancels the anomaly of the gauged $U(1)$ axial symmetry,
 is identified with the
$\eta '$ meson. In a particular set-up of
D3-branes on a $\c3z3$ orbifold singularity, the $\eta '$ meson is a
 twisted-sector R-R field.

\end{abstract}

\vspace{3cm}

\end{titlepage}

\section{Introduction}

\noindent

The $U(1)$ problem was considered as one of the major issues in
particle physics in the $70$'s. The dynamical breaking of the
$U_{\rm L}(3)\times U_{\rm R}(3)$ chiral symmetry to the diagonal
$U_{\rm V}(3)$ should
result in nine Goldstone
bosons, whereas only a light octet is observed in nature. The $\eta'$ meson
that correspond to the breaking of the $U_{\rm A}(1)$ symmetry is too heavy to be a
Goldstone.

It was understood that, at the qualitative level, the resolution of the
puzzle should involve the ABJ anomaly. It was later suggested by 't Hooft that the
$\eta '$ meson becomes massive because of instantons \cite{thooft}. A different
solution, which will be reviewed here, was suggested by Witten
\cite{Witten:1979vv} and by Veneziano \cite{Veneziano:1979ec}. They 
showed that the $\eta '$ should be massless in the 't Hooft large-$N$
limit (the planar theory), since the anomaly is a $1/N$ effect, and that a $1/N$ mass
for the $\eta '$ is generated when the anomaly is taken into account.
Their analysis resulted in the celebrated ``Witten--Veneziano
formula'' (WV) \cite{Witten:1979vv,Veneziano:1979ec}:
\beq
M_{\eta '} ^2 = {4 N_f \over f _ {\pi} ^2} {1\over
(16\pi ^2)^2} \int d^4 x \, \langle {\rm tr}\, F \tilde F (x)\, {\rm tr}\, F \tilde F (0)
\rangle |_{N_f=0}\,\,\, .
\label{WV}
\eeq

\begin{figure}
\epsfxsize=15cm
\centerline{\epsfbox{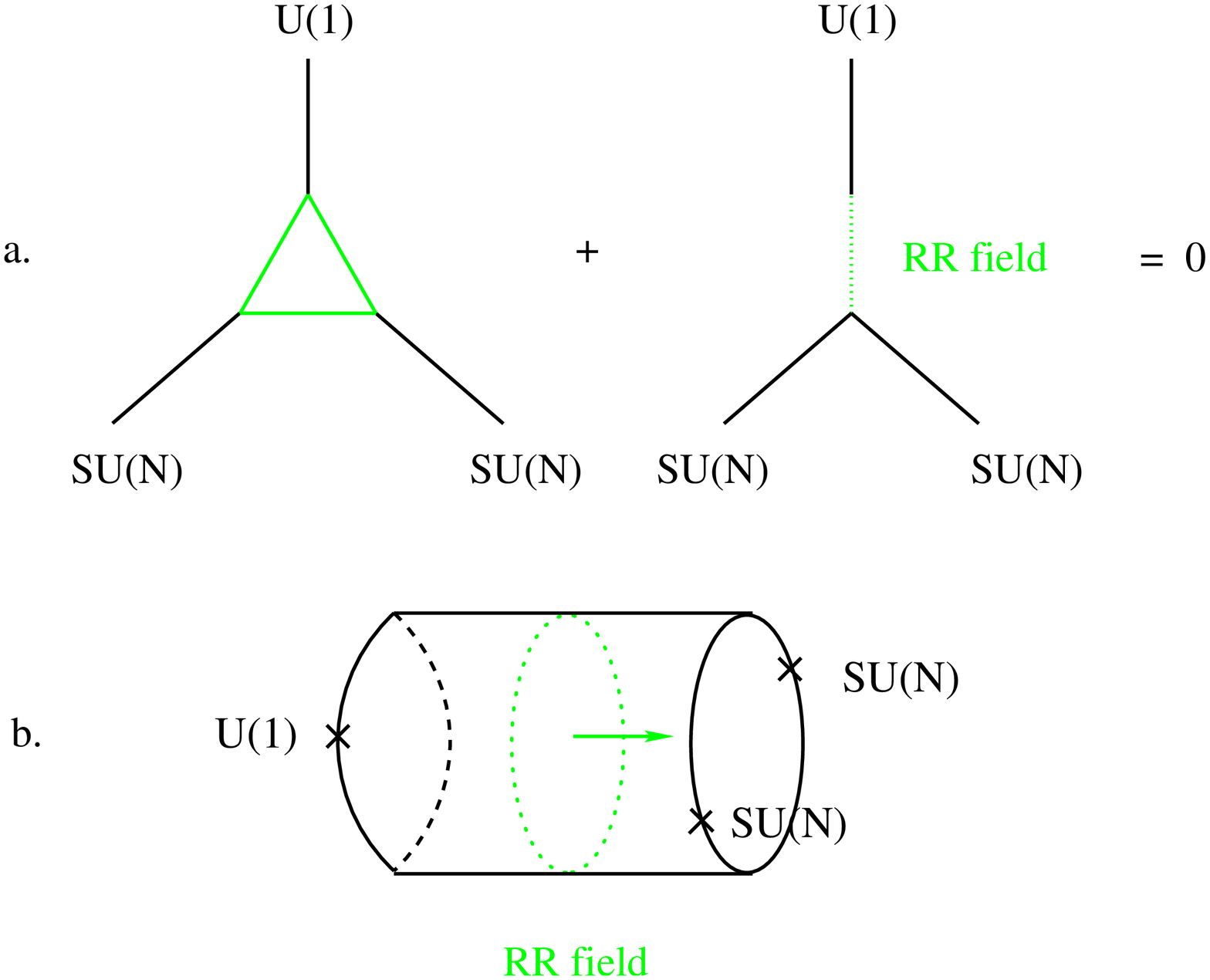}}
\caption{
a. The Green--Schwarz mechanism. b. The string theory diagram: the R-R
closed string mode is identified with the $\eta '$ meson.}
\label{GSfig}
\end{figure}

In this note we would like to derive the WV formula \eqref{WV} from the AdS/CFT
correspondence \cite{AdSCFT}. Aspects of chiral symmetry breaking
within the \newline ``AdS/CFT with flavor'' approach were discussed recently in
\cite{AdSflavor}\footnote{Note that in those models the $\eta '$ is
massless, since the limit $N_f/N_c \rightarrow 0$ is assumed.}. As we shall see the derivation links the
WV formula to the Green--Schwarz (GS) mechanism \cite{Green:sg}: the closed string
that is needed to cancel the $U_A(1)$ anomaly will be identified with the
$\eta '$ meson, see fig. \ref{GSfig}.

The set-up that we will use is D3-branes on orbifold singularities.
Although we will use a specific orbifold, we suggest that this phenomenon
is common to a generic set-up of D-branes on orbifold singularities. In
particular it does not require supersymmetry. In fact, it should be
even more general: 
in string theory there are no global symmetries. If
the set-up involves the field-theory $U(1)$ axial symmetry as a
symmetry of string theory\footnote{The $U(1)$ axial symmetry might be
 an ``accidental symmetry'' as well.}, it must be local. Local
symmetries cannot be anomalous and hence the anomaly must be canceled
via a generalized GS
mechanism. The intermediate (odd parity R-R) closed string mode, which
is involved in the anomaly cancellation, is the $\eta '$ meson. Another remark is
that we are taking the standard AdS/CFT decoupling limit: $\alpha'
\rightarrow 0$, while keeping $R^2 / \alpha '$ fixed. The proposed GS
mechanism ``survives'' this limit, as the anomaly is a field theory
effect. 

While our analysis is general, it would be interesting to make an
explicit calculation of the $\eta '$ mass in the resolved $\c3z3$
orbifold model (in units of the string tension).

The rest of this note is organized as follows: in section 2 we describe our
model and in section 3 we show how, within our specific set-up, 
we can derive the WV formula.

\section{D3 branes on $\c3z3$ orbifold \newline singularity}

\noindent

In this section we set the general framework. Consider, as in
\cite{Kachru:1998ys}, the type IIB string with 
a stack of $N$ D3-branes placed on a $\cgamma$ orbifold
singularity. In particular we will be interested in the $\c3z3$
orbifold. This model, as well as the general case of D3-branes on
$\ckk$ orbifold singularities, were analyzed in \cite{Hanany:1998it}.

The conformal field theory on the D3-branes is a supersymmetric chiral
$U^9(N)$ gauge theory with bi-fundamental matter.
Therefore, there are mixed anomalies of the form $U_i(1) SU^2_j(N)$
 (in fact, some combinations of $U(1)$'s are anomaly-free).

 The resolution of this local anomaly problem in the $\c3z3$ model is
via a generalized GS mechanism \cite{Ibanez:1998qp}. The
anomalous $U(1)$'s receive a mass of the string scale and therefore
they are infinitely massive from the field theory point of view. The low-energy
gauge group is thus $SU^9(N)$ (plus additional anomaly-free $U(1)$'s). In addition the local
anomalies are canceled via an exchange of massless closed strings, see
fig. \ref{GSfig}. In
 the above model the
closed strings modes are twisted-sector R-R fields. The terms that
give mass to the $U(1)$'s and cancel the anomaly are WZ-terms, which
are localized on the world-volume of the D3-branes:
\beq
\int d^4 x\ C \wedge \exp F\, .
\eeq

Indeed, the result of the detailed analysis
of ref.\cite{Hanany:1998it} is that there are six R-R (and NS-NS) 
twisted-sector moduli. They live on ${\mathbb C} \in {\mathbb C}^3 $,
hence they are six-dimensional.

In the AdS/CFT framework (the near-horizon limit of the
configuration), we consider type IIB string theory on $AdS_5 \times
S^5 / ({\mathbb Z}_3 \times{\mathbb Z}_3)$. The R-R twisted
sector moduli that cancel that $U(1)$ anomalies live on $AdS_5 \times S^1$.

The above model is satisfactory for most of our purposes; however, in
order to have a clearer picture, we would still like to make two
simplifications. We wish to turn off the coupling of all but one local
$U(N_c)$ group. In the string theory set-up, combinations of gauge couplings are controlled by NS-NS
moduli. By varying the six-dimensional twisted NS-NS moduli, we can
achieve the limit where the coupling of eight of the nine gauge
factors is much smaller than the coupling of the ninth. This is
clearly seen in the T-dual picture \cite{Hanany:1998it}.  

The second simplification that we wish to make is to change the rank
of the symmetry group so that
we will have a $U_{\rm L}(n_f)\times U(N_c) \times U_{\rm R}(n_f)$ group. String
theory allows us to do so by adding fractional branes at the orbifold
singularity. A different way of achieving an arbitrary number of
flavors, including the situation with no flavors at all, is to start
with the T-dual picture of D5-branes and NS5-branes and to vary the
number of D5-branes \cite{Hanany:1998it}. Thus we will consider the model in table \ref{content2}.

\begin{table}
\begin{displaymath}
\begin{array}{l c@{ } c@{ } c@{ } c}
 & \multicolumn{1}{c@{\times}}{U_{\rm L}(n_f)}
& \multicolumn{1}{c@{\times}}{U(N_c)}
& \multicolumn{1}{c@ {}}{U_{\rm R}(n_f)} \\
\hline
3\ \rm {chirals} & \Yfund & \overline{\Yfund}  & 1 \\
3\ \rm {chirals} & 1 & \Yfund & \overline{\Yfund}  \\
\end{array}
\end{displaymath}
\caption{The simplified model: a $U(N_c)$ gauge theory with
fundamental and anti-fundamental matter.}
\label{content2}
\end{table} 

Since the two groups $U_{\rm L}(1)$ and $U_{\rm R}(1)$ are global (or very weakly coupled), the
twisted-sector R-R fields will not lift their masses. Only the mass of the center of the
$U(N_c)$, which can be identified with the $U_{\rm B}(1)$ baryon number, is
lifted by the mixed anomalies $SU^2_{\rm L}(n_f) U_{\rm B}(1)$ and
$SU^2_{\rm R}(n_f) U_{\rm B}(1)$.
Thus the model we consider here is a $SU(N_c)$ gauge theory with $N_f \equiv 3n_f$ fundamental fermions (and scalars) and $N_f$
anti-fundamentals fields. Actually, when all (but the gauge)
interactions are turned off, the global symmetry is enhanced to $SU(N_f)$.
The GS action, in terms of two R-R fields,
 denoted by $C_{\rm L}$ and $C_{\rm R}$, takes the following form (see ref.
\cite{Armoni:2002fh} for an explicit derivation)

\newpage

\bea 
& &  
S = \int d^4 x\   {1\over \alpha'} \left ( {1\over 2} (\partial C_{\rm
L} - \partial C_{\rm R})^2 + {1\over 2 } (\partial C_{\rm L} + \partial
C_{\rm R} - {\rm tr}\, A )^2 \right )  \nonumber \\
& & + \int d^4 x\  {N_f \over 8\pi ^2}  (C_{\rm L} - C_{\rm R})
{\rm tr}\, F\tilde F .
 \label{GSaction2}
\eea

Note that the anomaly is proportional to $N_f$. The action $C_{\rm L}
\rightarrow C_{\rm L} + \alpha _{\rm L} $ ($C_{\rm R} \rightarrow
C_{\rm R} + \alpha _ {\rm R}$)
corresponds to a chiral rotation of the fundamental (anti-fundamental)
fermions. A somewhat simpler way of writing \eqref{GSaction2} is in
terms of the combinations $C_{\rm A} \equiv C_{\rm L} - C_{\rm R}$ and
$C_{\rm V} \equiv C_{\rm L}+C_{\rm R}$
\beq 
S = \int d^4 x\   {1\over \alpha'} \left ( {1\over 2} (\partial C_{\rm
A})^2 + {1\over 2 } (\partial C_{\rm V} - {\rm tr}\, A )^2 \right ) +
{N_f\over 8\pi ^2} C_{\rm A}
{\rm tr}\, F\tilde F .
 \label{GSaction3}
\eeq
The symmetry $C_{\rm A} \rightarrow C_{\rm A} + \alpha $, which became global
once the gauge couplings of the $U_{\rm L}(n_f),U_{\rm R}(n_f)$ gauge groups were
turned off, is the axial symmetry. Note the similarity of \eqref{GSaction3}
 to the QCD effective action for the $\eta '$ \cite{eff} (see also
\cite{Kawarabayashi:1980dp}), upon the
identification $C_{\rm A} = -\eta '$.

\section{A derivation of the Witten--Veneziano \newline formula}

\noindent

We now wish to derive the WV formula \eqref{WV} from the AdS/CFT
correspondence. In the AdS/CFT framework, color singlets are described
by closed strings \cite{Witten:1998zw}. We suggest that the $\eta '$ meson of the $\c3z3$
model, namely the lightest pseudo-scalar meson, is the twisted R-R
field $C_{\rm A}$. The arguments in favor of this suggestion are
clear: the action $C_{\rm A}
\rightarrow C_{\rm A} +\alpha$ is the generator of the axial symmetry.
Moreover, the field
$C_{\rm A}$ is a pseudo-scalar and it couples to the anomaly with a
strength $N_f/N_c$, as we expect from the $\eta '$ \footnote{In our
set-up, the ratio
$N_f/N_c$ is kept fixed, while $N_c\rightarrow \infty$; the $\eta
'$ is therefore expected to have a non-vanishing mass (in
contrast to the set-ups in refs.\cite{AdSflavor}).}.
Finally, we will show below
that the mass of the $C_{\rm A}$ field admits the WV formula \eqref{WV}.

We start by an evaluation of the two-point function $\langle {\rm tr}\, F\tilde
F(x)\, {\rm tr}\, F\tilde F(0)\rangle$. The AdS/CFT prescription for calculating
two-point functions $\langle O(x)\, O(0) \rangle$ in the boundary
field theory is to evaluate the propagators
of the bulk closed strings that couple to the boundary operator $O(x)$
\cite{AdSCFT}.
There are two bulk fields that couple to ${\rm tr}\, F\tilde F$:
the twisted sector R-R field $C_{\rm A}$ \eqref{GSaction2} and the ten
dimensional R-R 0-form $C$ (the ``axion''). In fact, it is well known
that the axion describes pseudo-scalar glueballs \cite{Csaki:1998qr}. Thus,
according to the AdS/CFT correspondence
\beq
 {1\over (16\pi ^2)^2} \langle {\rm tr}\, F \tilde F (x) \, {\rm tr}\, F \tilde F (0) 
\rangle = \xi ^2 \sum _ {\rm C\ modes} K_C (x,0) + \lambda ^2 \sum
_{\rm C_A\ modes} K_{C_{\rm A}} (x,0)\, ,
\label{twopoint}
\eeq
where $\xi,\lambda$ are the couplings of the ten-dimensional R-R 0-form
and the twisted sector R-R 0-form to $O= {\rm tr} F\tilde F$, respectively.
$K(x,0)$ is a boundary-to-boundary propagator. In the following we
will identify the $C$-modes and the $C_{\rm A}$-modes with glueballs
and flavor singlet mesons respectively. 

 The propagator of a
massless bulk field in a {\em confining} supergravity background is the
same as that of a free
massive field in a flat space \cite{Witten:1998zw}.
It is seen by
taking the ansatz $K(x;r) = e ^{iqx} F(r)$ and by solving the 
5-d bulk wave equation. Indeed, for a background of the form
\beq
ds^2 = dr^2 + f(r) dx_\mu ^2 \, ,
\eeq
 The bulk wave-equation takes the form
\beq
f^{-1}(r) \partial _r \left ( f^2(r) \partial _r F(r) \right ) -q ^2 F(r)=0 .
\label{bulk}
\eeq 
The eigenvalues $q^2$ are interpreted as the 4-d mass of the
color-singlet hadron and thus the boundary-to-boundary propagators, in
momentum space, take the form 
\beq
K_C  = {1\over q^2 - M_C ^2} \,\,\,\,\, ; \,\,\,\,\,
K_{C_{\rm A}} = {1\over q^2 - M_{C_{\rm A}} ^2} .
\label{props}
\eeq
Since the theory contains massless quarks \footnote{The gauginos, on
the other hand, are
assumed to have a small mass. It can easily be realized in the AdS/CFT
framework by an appropriate perturbation of the super-gravity solution.} 
\beq
\int d^4 x\, \langle {\rm tr}\, F \tilde F (x) \, {\rm tr}\, F \tilde
F (0) \rangle = 0 \, .
\eeq
(There is no meaning to a theta vacuum, since we can use a
chiral rotation to set the value of theta to an arbitrary value). Let
us proceed as in \cite{Witten:1979vv}. In the version of our theory without matter,
namely when $N_f/N_c =0$, the quantity $\int d^4 x \, \langle {\rm tr}\, F \tilde F
(x) \, {\rm tr}\, F \tilde F (0) \rangle $ is non-vanishing. We therefore must
conclude that in eq.\eqref{twopoint} the contribution from the mesons
cancels that from the glueballs: 
\beq
{1\over (16\pi ^2)^2}
\int d^4 x\, e^ {iqx} \, \langle {\rm tr}\, F \tilde F (x) \, {\rm tr}\, F \tilde F (0) \rangle
|_{{N_f\over N_c}=0,q^2=0} = {\lambda ^2 \over M _{C_A} ^2} 
\label{part1WV}
\eeq

In fact, when $N_f/N_c \rightarrow 0$ the sum over mesons ($C_{\rm
A}$-modes) is dominated by the lightest mode whose mass tends to zero
since in this limit the anomaly vanishes and $C_{\rm
A}$ becomes a true Goldstone. In this limit $C_{\rm A}$ does not
couple to the boundary field theory and \eqref{bulk} is solved by
$F(r)={\rm const.}$.

The second ingredient that is needed to complete our derivation is
the anomaly equation. From \eqref{GSaction3} we can read the equation
of motion for $C_{\rm A}$
\beq
{1\over \alpha '} \partial _\mu \partial ^\mu C_{\rm A}= {N_f \over 8\pi^2}
 {\rm tr}\, F \tilde F \, .
\label{anomaly}
\eeq
We therefore identify ${1\over \alpha '}\partial ^\mu C_{\rm A}$ with the axial current $J^\mu$. The
coupling of the r.h.s. of \eqref{anomaly} to $C_{\rm A}$ is by definition
$2N_f \lambda$. Following ref.\cite{Witten:1979vv} we argue that the coupling of 
$\partial _\mu J^\mu$ to $C_{\rm A}$ is proportional to $M^2_{C_{\rm A}}$, with a
proportionality parameter $f_{\rm A}$. The reason is that, in momentum
space, the coupling of $J^\mu$ should be $\sim q^\mu$ (the momentum is
the only vector in the problem) and therefore the coupling of $q_\mu
J^\mu$ is $\sim q^2 = M^2_{C_{\rm A}}$. Thus
\beq
 2N_f \lambda =  f_{\rm A} M^2_{C_{\rm A}}\,.
\label{part2WV}
\eeq
Inserting \eqref{part2WV} in \eqref{part1WV}
\beq
M^2_{C_{\rm A}} ={4N_f ^2 \over f_{\rm A}^2} 
{1\over (16\pi ^2)^2}
\int d^4 x\, e^ {iqx} \, \langle {\rm tr} \, F \tilde F (x) \, {\rm tr}
\, F \tilde F (0) \rangle
|_{{N_f\over N_c}=0,q^2=0}\, ,
\eeq
we arrive at the WV formula \eqref{WV},
provided that $C_{\rm A}$ is identified with $\eta'$ and its coupling square
$f^2 _{\rm A}$ with $N_f f^2 _{\pi }$.   

{\it Note added:} I have been informed by J. Barbon of a related
work \cite{pepe}, where the WV formula is discussed within the
framework of ``AdS/CFT with flavor''.

{\bf Acknowledgements:}
I would like to thank J. Barbon, C. Hoyos, E. Lopez, A. Ritz and
G. Veneziano for discussions and for comments on the manuscript.

\end{document}